\documentclass[a4paper,11pt]{article}
\pdfoutput=1 

\usepackage{jinstpub} 
\usepackage{here}
\usepackage{comment}
\usepackage[natbib=false,style=numeric,sorting=none]{biblatex}
\makeatletter
\newcommand{\figcaption}[1]{\def\@captype{figure}\caption{#1}}
\newcommand{\tblcaption}[1]{\def\@captype{table}\caption{#1}}
\makeatother

\title{\boldmath Analysis of SiW-ECAL technological prototype beam test with electron beam }

\author[a,1]{Y. Kato,\note{Corresponding author.}}
\author[b]{K. Goto,}
\author[b]{and T. Suehara}

\affiliation[a]{The University of Tokyo,\\7-3-1 Hongo, Bunkyo-ku, Tokyo, Japan}
\affiliation[b]{Kyushu University,\\744 Motooka, Nishi-ku,
Fukuoka, Japan}

\emailAdd{katou@icepp.s.u-tokyo.ac.jp}

\abstract{
 The beam test for the SiW-ECAL technological prototype was conducted using electron beam in June-July 2019 at DESY test beam facility in Hamburg, Germany.
 In the beam test, there were two main programs: MIP program (without tungsten) and shower program (with tungsten). 
 The purposes of the MIP program are energy calibration, pedestal uniformity/stability assessment, TDC operation test, and retrigger verification.
 On the other hand, the shower program aims to evaluate energy resolution through beam energy measurement. 
 In addition, several operation tests with FEV13 were also conducted such as individual threshold control and auto gain mode.
 This article presents the performance of the technological prototype of SiW-ECAL, mainly focusing on the version of FEV13-Jp.
}

\keywords{Calorimeters, Calorimeter methods, Si microstrip and pad detectors, Performance of High Energy Physics Detectors}

\collaboration[c]{on behalf of CALICE SiW-ECAL group}

\proceeding{CHEF2019 \\
  25-29 November 2019\\
  Centennial Hall, Kyushu University}

\begin{document}
\maketitle
\flushbottom


\section{Introduction}

The next generation Higgs Factory, such as the International Linear Collider (ILC) project~\cite{TDR1}, can be a powerful approach to new physics based on Higgs precision measurements. 
The International Large Detector (ILD)~\cite{TDR4} which is one of the detector concept for the ILC is designed to be optimized for the Particle Flow Algorithm (PFA)~\cite{Thomson_2009} to enable the most sensitive measurement of Higgs.
The PFA is expected to improve basic particle level reconstruction and jet energy resolution by particle identification and measuring momentum of charged tracks and energy of neutral clusters separately.
To realize good separation of hadronic decays from W and Z, the PFA requires a high granularity calorimeter system, specifically cell size of around 5 $\times$ 5 mm$^2$ for electromagnetic calorimeter (ECAL) and 30 $\times$ 30 mm$^2$ for hadronic calorimeter (HCAL).

The ILD ECAL is designed as a sampling type calorimeter consisting of 30 layers of sensors and tungsten absorbers with 24 radiation lengths. 
There are two types of sensor option, one is silicon pixel sensor and the other is scintillator with photo sensor.
We are researching and developing the silicon-tungsten (SiW) ECAL within the framework of CALICE collaboration, which aims to develop highly granular calorimeters to be operated at future colliders.

This document reports on the performance of the technological prototype of SiW-ECAL, mainly focusing on the version of FEV13-Jp, in electron beam test carried out at DESY in June - July 2019.

\section{Development status of SiW-ECAL: Technological prototype}

The first prototype of SiW-ECAL was the so-called ``Physics prototype''~\cite{Adloff:2008aa} which proved the concept of particle flow calorimetry with several beam tests in 2005-2011.
Whereas the physics prototype was not optimized with respect to readout system and power consumption etc., the next generation prototype called ``Technological prototype'' has been developed since 2011 to establish the whole system of real ILD ECAL.
There are several technical challenges, such as:
Embedded electronics with dedicated ASICs in 1-3 mm thick electronics layers,
Smaller cell size of around 5 $\times$ 5 mm$^2$ (10 $\times$ 10 mm$^2$ in the physics prototype),
Power pulsing operation synchronized with the ILC bunch structure,
Buffering up to 15 event frames in single bunch with self-triggering capability, and
Multiple ASICs connected on the bus structure.

The technological prototype of SiW-ECAL principally consists of Active Sensor Unit (ASU) and control/readout interface.
The ASU is a set of silicon sensors, printed circuit board (PCB) and application specific integrated circuits (ASICs) and its size is 18 $\times$ 18 cm$^2$.
A silicon wafer has $16 \times 16 = 256$ sensors which pixel size is  5.5 $\times$ 5.5 mm$^2$.
The size of wafer is 9 $\times$ 9 cm$^2$ therefore 4 wafers are mounted on the single ASU and the ASU has totally 1024 channels.
There are several options of silicon thickness: 320, 500, 650 $\mathrm{\mu}$m.
The dedicated ASIC is designed for SiW-ECAL and developed by Omega group in France, called SKIROC (Silicon Kalorimeter ReadOut Chip)\cite{Callier_2011} which can handle 64 channels.
It has self-triggering capability, dual gain slow shapers for charge measurement, 15 analog memories made of switched capacitor array (SCA), 12-bit Wilkinson ADC and 12 bit counter for tagging bunch crossing ID (BCID) of each event in a ILC train.
BCID is used for coarse timing measurement which resolution is 0.2 $\mathrm{\mu s}$.
In addition, SKIROC has TDC capability with fine timing resolution of a few ns.
Although only two of high/low gain or TDC output can be read at once, SKIROC has a function to select a suitable gain of charge output automatically.
Please refer to \cite{Callier_2011} for details about SKIROC.

\begin{figure}[H]
    \centering
    \begin{tabular}{cc}
    \begin{minipage}{0.48\hsize}
    \centering
    \includegraphics[width=7truecm]{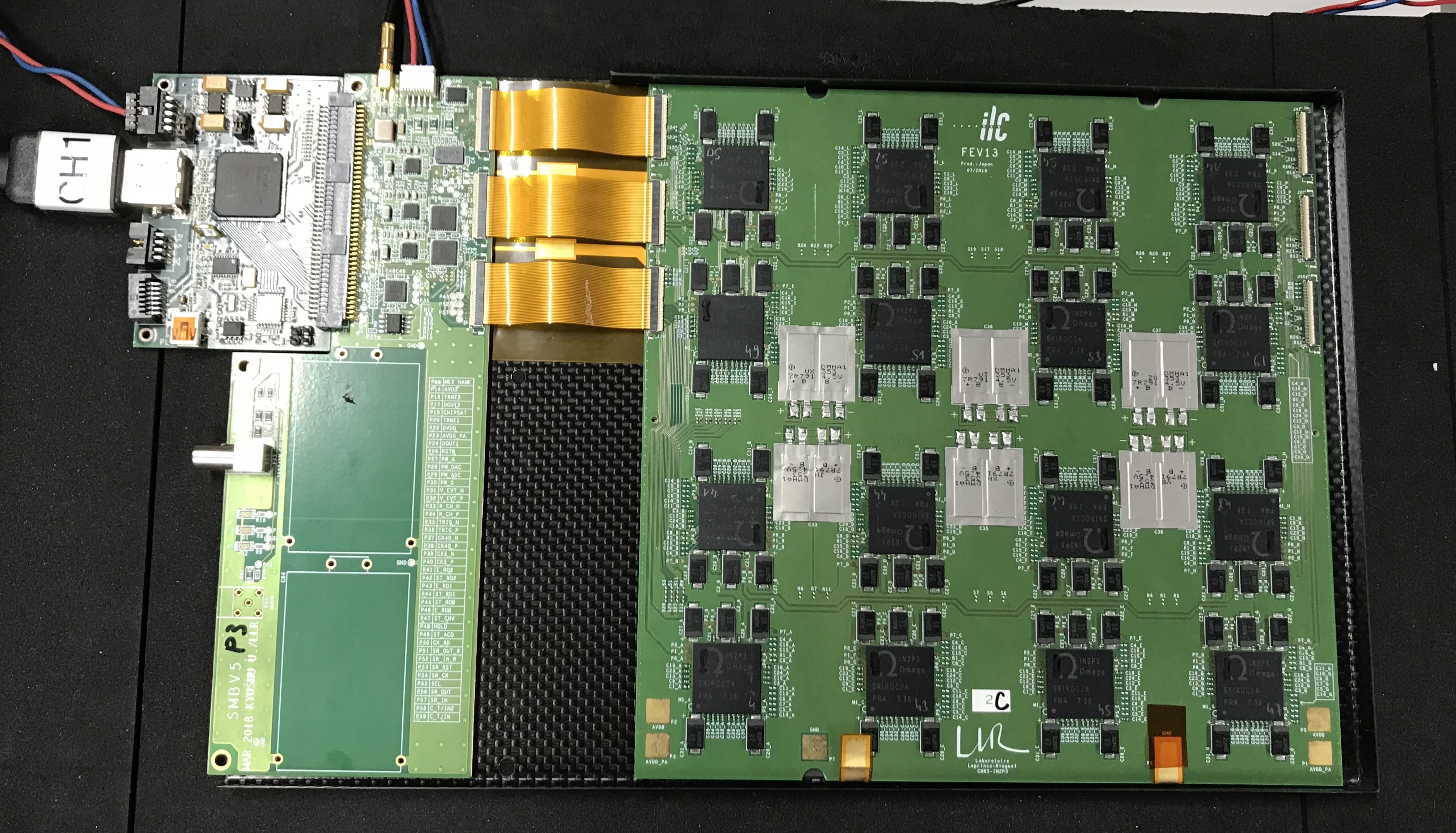}    
    \end{minipage}
    \begin{minipage}{0.48\hsize}
    \centering
    \includegraphics[width=7truecm]{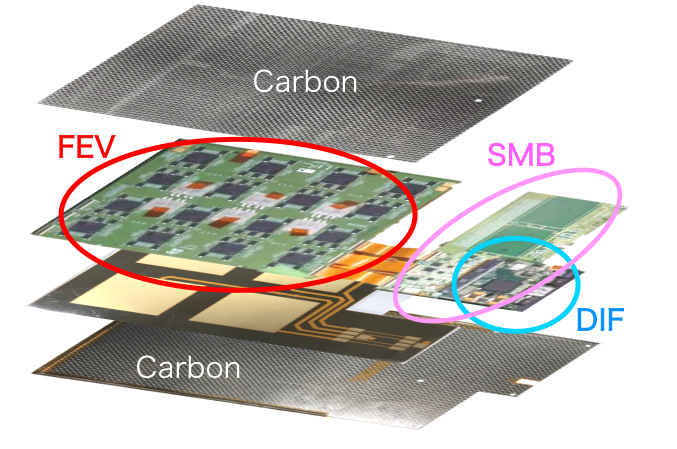}    
    \end{minipage}
    \end{tabular}
    \caption{Structure of FEV13-Jp.}
    \label{fig:fev13}
\end{figure}

Figure~\ref{fig:fev13} shows the current prototype we are developing called “FEV13-Jp”. 
It is equipped with four silicon wafers which thickness is 320 or 650 $\mathrm{\mu}$m and 16 ASICs packaged in BGA. 
It also has a kapton film to apply bias voltage (150-180 V) to sensors, cooling and electric shielding made of carbon fiber sheet, detector interface (DIF) to send data to the DAQ system, adaptor board called SMB between the DIF and the ASU providing support for voltage supplies, and a big capacitance for power pulsing.
Alternative Chip-On-Board (COB) implementation of the ASICs and new control/readout interface called SL-Board (SLB) is also being studied by LLR and LAL.
The current version of SKIROC is 2A~\cite{Suehara_2017}, which expands the range of individual threshold adjustment for each channel, improves the signal-over-noise (S/N) ratio for both of triggers and ADCs, and also TDC resolutions.

\section{Electron Beam Test at DESY}

\begin{figure}[H]
    \centering
    \begin{tabular}{cc}
    \begin{minipage}{0.38\hsize}
    \centering
    \includegraphics[width=7truecm]{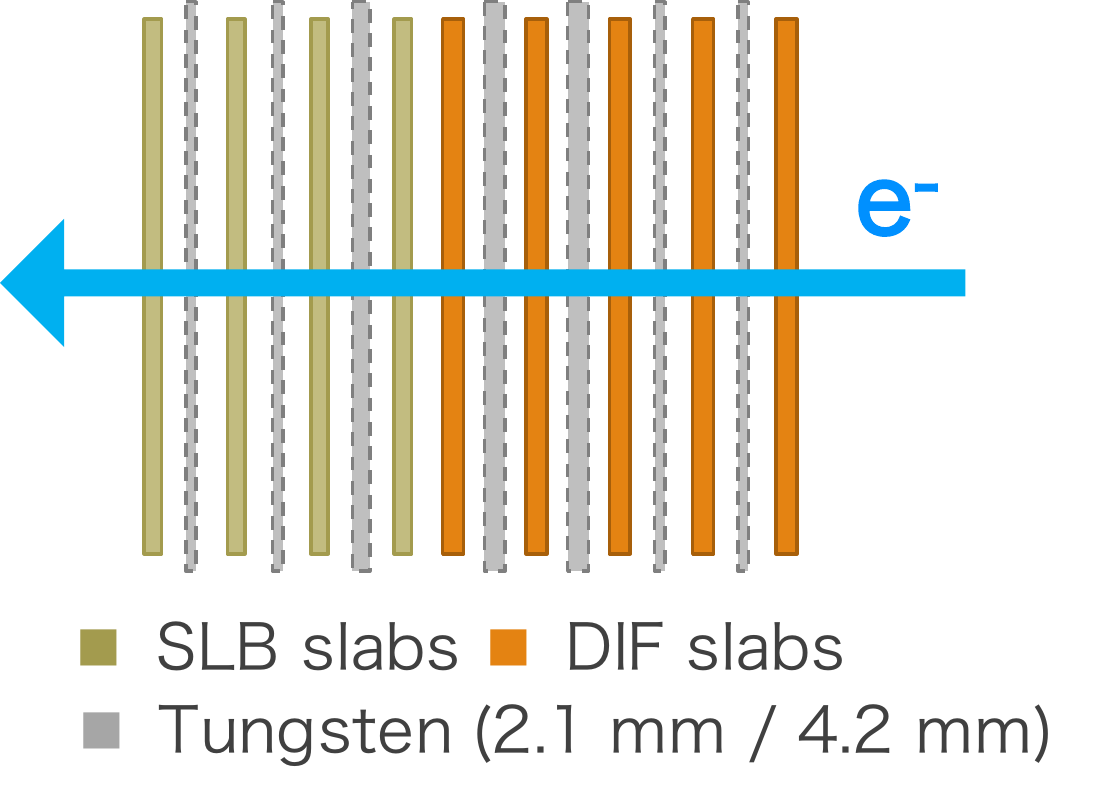}    
    \end{minipage}
    \begin{minipage}{0.58\hsize}
    \centering
    \includegraphics[width=7truecm]{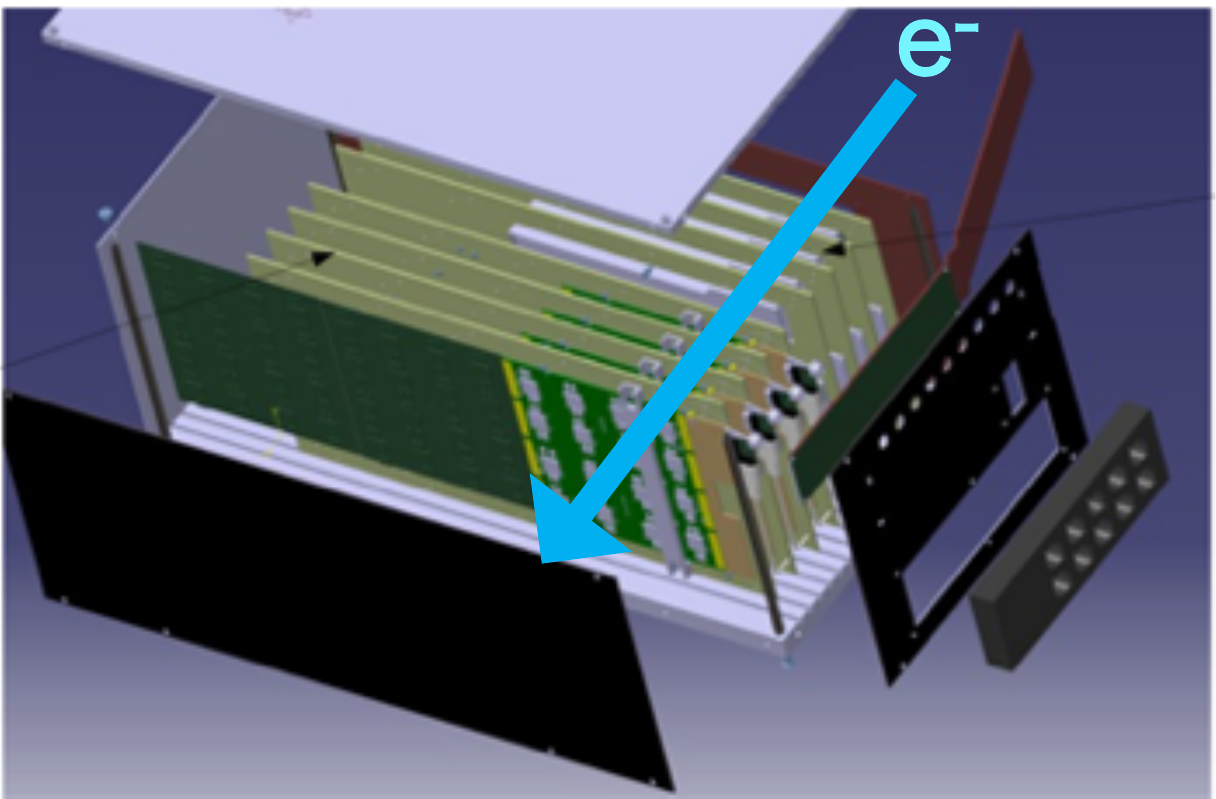}    
    \end{minipage}
    \end{tabular}
    \caption{Setup of the beam test at DESY.}
    \label{fig:BTsetups}
\end{figure}

Beam test for SiW-ECAL technological prototype was carried out using 1-5 GeV electron beam at DESY test beam facility (TB24) in June-July 2019.
All the slabs tested during this beam time are summarized in Tab.~\ref{tab:slabs}.
Figure~\ref{fig:BTsetups} shows schematics of the beam test setup.
The interval between slabs is 30 mm and we can install/remove tungsten layer.

\begin{table}[H]
\centering
\caption{List of slabs tested during this beam time.}
\begin{tabular}{|c|ccccc|cccc|}
\hline
slab ID                   & P1    & P2    & P3    & K1    & K2    & SL1.0 & SL1.1 & SL1.2 & SL1.3 \\ 
\hline
interface                 & DIF   & DIF   & DIF   & DIF   & DIF   & SLB   & SLB   & SLB   & SLB   \\ 
PCB                       & FEV13 & FEV13 & FEV13 & FEV13 & FEV13 & COB   & FEV12 & COB   & FEV12 \\ 
Si {[}$\mu$m{]} & 650   & 650   & 320   & 650   & 650   & 500   & 500   & 500   & 500   \\ \hline
\end{tabular}
\label{tab:slabs}
\end{table}

There are principally three objectives of this beam test: comparison of ASU based on BGA (FEV) and based on COB, test of SLB, and validation of FEV13-Jp which is the target of this document. 
The physics program of the beam test can be summarized in the following points:
\begin{itemize}
    \setlength{\parskip}{0.05cm}
    \setlength{\itemsep}{0.01cm}
    \item Commissioning and calibration using 3 GeV electrons acting as minimum ionizing particle (MIPs) without absorber layer
    \item Measurement of response from EM shower with tungsten absorber with 1-5 GeV electrons.
\end{itemize}

\section{Performance of FEV13-Jp}

\subsection{Trigger adjustment}

The commissioning of the slabs were performed for trigger adjustment before the beam test.
The threshold of the internal trigger for each ASIC can be optimized by scanning of trigger threshold with measuring the noise rate of all the channels.
In this beam time, we did not use the individual adjustment of threshold by 4-bit DAC and masked the significantly noisy channels.
As a result, we decided the trigger threshold of 240 DAC for all the ASICs and masked at most one or two channels in one ASIC.

\begin{figure}[h]
  \begin{minipage}[c]{.48\textwidth}
     \includegraphics[width=8truecm]{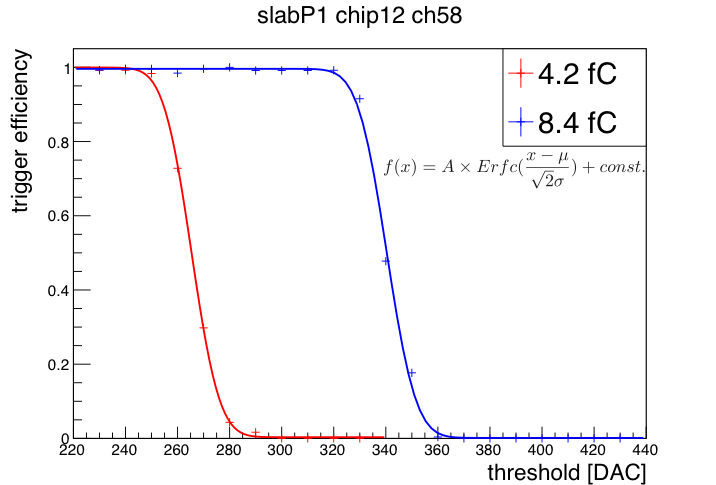}   
    \caption{S-curve for slab P1, chip 12, channel 58.}
    \label{fig:scv}
  \end{minipage}
  \hfill
  \def\@captype{table}
  \begin{minipage}[b]{.48\textwidth}
    \tblcaption{Summary of S/N ratio for the trigger with channel 56-63 in all the ASICs of slab P1.}
    \begin{tabular}{|c|c|c|}
    \hline
    injection [fC] & 4.2              & 8.4             \\ \hline
    $\mu$ [DAC]    & 261.9 $\pm$ 3.0  & 334.3 $\pm$ 3.9 \\
    $\sigma$ [DAC] & 8.0 $\pm$ 0.8    & 8.1 $\pm$ 0.9   \\ \hline
    S/N$_{trig}$ & \multicolumn{2}{c|}{9.0 $\pm$ 1.1} \\ \hline
    \end{tabular}
    \label{tab:sntrg}
  \end{minipage}
\end{figure}

We also performed threshold scan with test injection for detailed evaluation of trigger performance after the beam time at laboratory in Kyushu University.
The injected charge of 4.2 fC is equivalent to response by 1 MIP for 320 $\mathrm{\mu}$m Si.
Figure~\ref{fig:scv} shows one of the trigger efficiency plots corresponding to each threshold called S-curve.
From these plots, we can estimate S/N ratio for the trigger defined as:
\begin{equation}
    S/N_{trig} \equiv \frac{\mu_{2MIP}-\mu_{1MIP}}{\sigma_{1MIP}}
\end{equation}
where $\mu$ and $\sigma$ are the parameters of complementary error function which is used to fit S-curves.
The results are summarized in Tab.~\ref{tab:sntrg}.

\subsection{Pedestal study}
\label{sec:ped}

Pedestal is the baseline of the charge output on the untriggered channel.
The uniformity and stability of the pedestals are essential for energy measurement.
After the removal of fake events\footnote{The fake event sometimes occur in which almost all the channel in certain ASIC are triggered in consecutive BCIDs and the pedestal may shift.}, we performed the fit of pedestal distribution with Gaussian for all the channels and memory cells.
The mean of pedestal is used for pedestal subtraction and width is for evaluation of S/N ratio for charge measurement.

We monitored pedestal mean value through the beam time and there were no significant fluctuation.
We also confirmed that pedestal width was around 3.5 which value is almost consistent with the previous tests.
However, we found that the pedestal of charge output in the TDC readout mode became noisier than only charge measurement mode.
The research for this behavior will be a higher priority task.

\subsection{MIP calibration}

For energy calibration of all the channels, we performed MIP run in which the 3 GeV electron beam was irradiated perpendicular to the slab plane.
After pedestal subtraction from charge output distributions, the ADC output value has linearity against the input charge.
The triggered spectrum of MIP is fitted by Laudau function convoluted by a Gaussian.
We can obtain the Most Probable Value (MPV) from the parameter of fit function and finally calculate S/N ratio for the charge measurement defined as:
\begin{equation}
    S/N_{charge} \equiv \frac{MPV_{1MIP}}{width_{pedestal}}.
\end{equation}

\begin{figure}[H]
    \centering
    \includegraphics[width=12truecm]{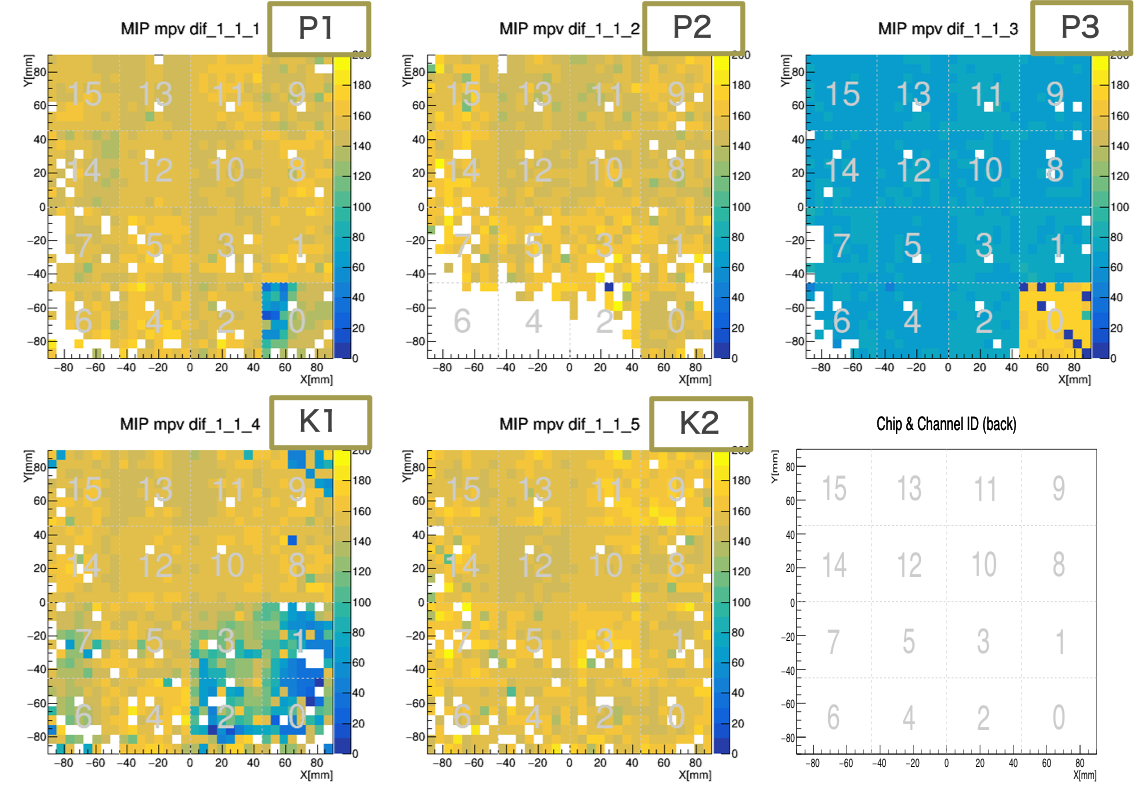}
    \caption{MPV distribution for all the FEV13-Jp.}
    \label{fig:mpv}
\end{figure}

Typical values of MPV, pedestal width and S/N ratio for charge are summarized for all the FEV13-Jp in Tab.~\ref{tab:snmip}.
In addition, Figure~\ref{fig:mpv} shows MPV distribution for all the channels.

\begin{table}[H]
\centering
\caption{Summary of typical S/N ratio for the charge measurement.}
\begin{tabular}{|c|ccccc|}
\hline
slab ID        & P1    & P2    & P3   & K1    & K2    \\ \hline
Si thickness [$\mu$m]   & 650   & 650   & 320  & 650   & 650   \\
MPV for 1MIP    & 146.5 & 144.9 & 71.3 & 141.4 & 146.1 \\
pedestal width & 3.0   & 3.0   & 3.3  & 2.8   & 3.1   \\
S/N$_{charge}$            & 49.0  & 48.9  & 21.7 & 50.2  & 47.5  \\ \hline
\end{tabular}
\label{tab:snmip}
\end{table}

Chip 0, 10-15 have enough statistics of MIP events to fit its spectrum with high accuracy.
However, chip 0 of slab P1 and P3 and chip 0-3 of slab K1 behave differently than others therefore they may have some problems in their ASICs or conductivity between PCB and silicon wafer.

\subsection{EM shower analysis}

The track is reconstructed by coincidence of BCIDs within $\pm 1$ between slabs.
After track reconstruction and energy calibration by MIP, we can finally measure the energy deposited by EM shower.
In the shower program, we installed tungsten layer as Fig.\ref{fig:BTsetups} where the total radiation length is 6.0.
As a first approach, the energy deposit in each channel is compared between FEV13-Jp slabs.
In addition, we compared them with simulated results using DDSim\cite{DD4hep} software.
This part is preliminary, therefore more detailed studies are in progress.

\subsection{TDC operation test}

Before the beam time, TDC calibration was performed  in laboratory at Kyushu University.
In order to scan the injection timing, a test pulse synchronized with the bunch clock was injected while shifting the phase.
As results, we obtained TDC output curve corresponding to injection timing and TDC calibration factors of 0.127 ns/TDC for up slope and 0.066 ns/TDC for down.
We also found that the phase is not match between TDC ramp and bunch clock, in addition, the TDC up slope may be saturated.
Further study should be done.

The TDC operation test was performed for the first time in the beam test.
In the MIP run (without absorber), TDC correlation between slabs were observed.
We also have to correct TDC time walk which depends on charge output.
In addition, as mentioned in Sec.~\ref{sec:ped}, pedestal problems were found during the TDC operation.
The further analysis is in progress.


\acknowledgments



The measurements leading to these results have been performed at the Test Beam Facility at DESY Hamburg (Germany), a member of the Helmholtz Association (HGF).





\printbibliography

@techreport{TDR1,
      author         = "Behnke, T. and Brau, James E. and Foster, Brian and
                        Fuster, Juan and Harrison, Mike and Paterson, James McEwan
                        and Peskin, Michael and Stanitzki, Marcel and Walker,
                        Nicholas and Yamamoto, Hitoshi",
      title          = "{The International Linear Collider Technical Design
                        Report - Volume 1: Executive Summary}",
      year           = "2013",
      eprint         = "1306.6327",
      archivePrefix  = "arXiv",
      primaryClass   = "physics.acc-ph",
      reportNumber   = "ILC-REPORT-2013-040, ANL-HEP-TR-13-20,
                        BNL-100603-2013-IR, IRFU-13-59, CERN-ATS-2013-037,
                        COCKCROFT-13-10, CLNS-13-2085, DESY-13-062,
                        FERMILAB-TM-2554, IHEP-AC-ILC-2013-001, INFN-13-04-LNF,
                        JAI-2013-001, JINR-E9-2013-35, JLAB-R-2013-01,
                        KEK-REPORT-2013-1, KNU-CHEP-ILC-2013-1, LLNL-TR-635539,
                        SLAC-R-1004, ILC-HIGRADE-REPORT-2013-003"
}

@techreport{TDR4,
      author         =  "Behnke, T. and Brau, James E. and Burrows, Philip N.
                        and Fuster, Juan and Peskin, Michael and Stanitzki, Marcel
                        and Sugimoto, Yasuhiro and Yamada, Sakue and Yamamoto,
                        Hitoshi and Abramowicz, Halina",
      title          = "{The International Linear Collider Technical Design
                        Report - Volume 4: Detectors}",
      year           = "2013",
      eprint         = "1306.6329",
      archivePrefix  = "arXiv",
      primaryClass   = "physics.ins-det",
      reportNumber   = "ILC-REPORT-2013-040, ANL-HEP-TR-13-20,
                        BNL-100603-2013-IR, IRFU-13-59, CERN-ATS-2013-037,
                        COCKCROFT-13-10, CLNS-13-2085, DESY-13-062,
                        FERMILAB-TM-2554, IHEP-AC-ILC-2013-001, INFN-13-04-LNF,
                        JAI-2013-001, JINR-E9-2013-35, JLAB-R-2013-01,
                        KEK-REPORT-2013-1, KNU-CHEP-ILC-2013-1, LLNL-TR-635539,
                        SLAC-R-1004, ILC-HIGRADE-REPORT-2013-003"
}

@misc{DD4hep,
author = {Frank, M. and Gaede, F. and Petric, Marko and Sailer, Andre},
title = {AIDASoft/DD4hep},
year = 2018,
doi = {10.5281/zenodo.592244},
url = {http://dd4hep.cern.ch/}
}

@article{Thomson_2009,
   title={Particle flow calorimetry and the PandoraPFA algorithm},
   volume={A611},
   DOI={10.1016/j.nima.2009.09.009},
   journal={Nucl. Instrum. Meth.},
   author={Thomson, M. A.},
   year={2009},
   pages={25-40},
   eprint = "0907.3577",
   archivePrefix  = "arXiv",
   primaryClass   = "physics.ins-det"
}

@article{Adloff:2008aa,
      author         = "Adloff, C. and others",
      title          = "{Response of the CALICE Si-W electromagnetic calorimeter
                        physics prototype to electrons}",
      collaboration  = "CALICE",
      journal        = "Nucl. Instrum. Meth.",
      volume         = "A608",
      year           = "2009",
      pages          = "372-383",
      doi            = "10.1016/j.nima.2009.07.026",
      eprint         = "0811.2354",
      archivePrefix  = "arXiv",
      primaryClass   = "physics.ins-det",
      reportNumber   = "CALICE Analysis Note CAN-008, CALICE-CAN-2008-002,
                        CALICE-PUB-2008-002"
}

@article{Callier_2011,
	doi = {10.1088/1748-0221/6/12/c12040},
	year = 2011,
	publisher = {{IOP} Publishing},
	volume = {6},
	pages = {C12040},
	author = {S. Callier and F. Dulucq and C de La Taille and G Martin-Chassard and N Seguin-Moreau},
	title = {{SKIROC}2, front end chip designed to readout the Electromagnetic {CALorimeter} at the {ILC}},
	journal = {JINST}
}

@article{Suehara_2017,
	doi = {10.1088/1742-6596/928/1/012039},
	year = 2017,
	publisher = {{IOP} Publishing},
	journal = {J. Phys.: Conf. Ser.},
	volume = {928},
	pages = {012039},
	author = {T. Suehara and others},
	title = {Towards a Technological Prototype for a High-granularity Electromagnetic Calorimeter for Future Lepton Colliders}
}
\end{document}